\documentclass[aps,twocolumn,showpacs,preprintnumbers,amsmath,amssymb,
nofootinbib,showkeys,floatfix]{revtex4-1}

\usepackage{epsfig}
\usepackage{graphicx}
\usepackage{dcolumn}
\usepackage{latexsym}
\usepackage{graphicx}
\usepackage{enumerate}
\usepackage{float}
\usepackage{placeins}
\usepackage{supertabular}
\usepackage{textcomp}

\usepackage{hyperref} 
\hypersetup{
   colorlinks=true,
   linkcolor=blue,
   citecolor=blue,
   urlcolor=blue
}

\def\tstrut{\vrule height2.5ex depth0pt width0pt} % used in tables

\bibpunct{[}{]}{,}{n}{,}{,} % options of natbib

\begin{document}

\title{Puzzles in hadronic transitions of heavy quarkonium with two pion emission}
\author{Jorge Segovia}\email{jsegovia@anl.gov}\affiliation{Physics Division, Argonne
National Laboratory, \\ Argonne, Illinois 60439-4832, USA}
\author{D.R. Entem}\email{entem@usal.es}\affiliation{Departamento de F\'{\i}sica
Fundamental and IUFFyM, \\ Universidad de Salamanca, E-37008 Salamanca, Spain}
\author{F. Fern\'andez}\email{fdz@usal.es}\affiliation{Departamento de F\'{\i}sica
Fundamental and IUFFyM, \\ Universidad de Salamanca, E-37008 Salamanca, Spain}
%
% \author{J. Segovia}%\email{segonza@usal.es}
% \author{D.R. Entem}%\email{entem@usal.es}
% \author{F. Fern\'andez}%\email{fdz@usal.es}
% %
% \affiliation{Grupo de F\'isica Nuclear and IUFFyM, \\ Universidad de 
% Salamanca, E-37008 Salamanca, Spain}
%
\date{\today}

\begin{abstract}
We study the anomalously large rates of some hadronic transitions observed in heavy
quarkonia using a constituent quark model which has been successful in describing
meson and baryon phenomenology. QCD multipole expansion (QCDME) is used to described
the hadronic transitions. The hybrid intermediate states needed in the QCDME method
are calculated in a natural, parameter-free extension of our constituent quark model
based on the Quark Confining String (QCS) scheme. Some of the anomalies are explained
due to the presence of a hybrid state with a mass near the one of the decaying
resonance whereas others are justified by the presence of molecular components in the
wave function. Certain unexpected results are pointed out.
\end{abstract}


\maketitle

\section{INTRODUCTION}
\label{sec:introduction}

The observation in the last decade of many new $c\bar{c}$ and charmonium-like states
has re-opened interest in charmonium spectroscopy. This interest has been also
extended to the bottomonium sector. Hadronic transitions of heavy quarkonia such as
$\psi(nS)$ or $\Upsilon(nS)$ to lower states with emission of two pions are important
means to study these new states and understanding both the heavy quarkonium dynamics
and the light hadron(s) formation. 

Many new results have been collected by the electron-positron colliders using the
initial state radiation technique. In particular, the $e^{+}e^{-}\to
J/\psi\pi^{+}\pi^{-}$ and $e^{+}e^{-}\to \psi(2S)\pi^{+}\pi^{-}$ reactions have been
recently analyzed in a mass distribution region between $3.5$ and $5.5\,{\rm
GeV/c^2}$. As the entrance channel fixes the quantum numbers $J^{PC}=1^{--}$, one
expects to obtain information of the possible vector charmonium resonances in this
energy region. However, the recent experimental data on hadronic transitions of
charmonium states which are above the open-flavor threshold show a puzzling
behavior. In the $J/\psi\pi^{+}\pi^{-}$ channel only one resonance appears,
attributed to the $X(4260)$, whereas in the $\psi(2S)\pi^{+}\pi^{-}$ channel two
resonances, compatible with the $X(4360)$ and $X(4660)$, show up. A recent
re-analysis of the $\psi(2S)\pi^{+}\pi^{-}$ data including the $X(4260)$ resonance
shows a non-significant contribution ($2.1\,\sigma$). No signals of the $\psi(4040)$,
$\psi(4160)$ and $\psi(4415)$ appears in the data. Furthermore, the $X(4360)$ and
$X(4660)$ resonances show an anomalous large width in the $\psi(2S)\pi^{+}\pi^{-}$
channel~\cite{Lees:2012cn,Wang:2007ea,Lees:2012pv}. In the bottom sector, compared to
the ordinary $\Upsilon(nS)\rightarrow \Upsilon(mS)$ $(m<n)$ transitions, the partial
widths of the $\Upsilon(10860)$ decaying into $\Upsilon(1S)$, $\Upsilon(2S)$ or
$\Upsilon(3S)$ plus two pions are out of line by two orders of
magnitude~\cite{Abe:2007tk}.

The anomalous widths can be due to several mechanisms: Contribution of hadron
loops~\cite{Meng:2007tk}; four-quark components in the quarkonium wave
functions~\cite{Ali:2010pq}; or, as we will see later, the existence of hybrid mesons
with a mass near the one of the decaying resonance.

Hadronic transitions can be described using the QCD multipole expansion (QCDME)
approach~\cite{Kuang:2006me}. In the single channel picture the light hadrons are
converted from the gluons emitted by the heavy quarks in the transition. The typical
momentum of the emitted gluons is too low for perturbative QCD to be applicable. Then
nonperturbative approaches, like QCDME, are needed.

In this approach the heavy quarkonium system serves as a compact color source which
emits two soft gluons that hadronize, for instance, into two pions. After the
emission of the first gluon and before the emission of the second one, there exists
an intermediate state where the $Q\bar{Q}$ pair together with the gluon forms a
hybrid state. The width of the transition depends critically on the position of this
state, therefore it is important to describe consistently the $Q\bar{Q}$ states and
the hybrids using as few parameters as possible. Apart from lattice
calculations~\cite{Juge:1999ie,Dudek:2008sz}, hybrid meson properties has been
calculated in different models: the flux-tube
model~\cite{Isgur:1984bm,Barnes:1995hc}, constituent gluons~\cite{Horn:1977rq},
Coulomb gauge QCD~\cite{Guo:2008yz} and quark confining string model
(QCS)~\cite{Tye:1975fz,Giles:1977mp,Buchmuller:1979gy} or QCD string
model~\cite{Kalashnikova:2008qr}.
 
In this work we will address the description of the new data of the hadronic
transitions in heavy quarkonium within the framework of a constituent quark model
(see references~\cite{Valcarce:2005em} and~\cite{Segovia:2013wma} for reviews) which
has been successful in describing the hadron phenomenology and the hadronic
reactions. Hybrid states are consistently generated in the original quark model using
the QCS scheme. In this way, we minimize the number of free parameters describing
both conventional and hybrid states.

The paper is organized as follows. In Sec.~\ref{sec:CQM} we will review the main
properties of the constituent quark model and give its prediction for the vector
charmonium and bottomonium states. Sec.~\ref{sec:QCDME} is devoted to the description
of the QCDME approach and the hybrid model we use. We will present our results in
Sec.~\ref{sec:results}. The work will be summarized in Sec.~\ref{sec:summary}.

\vspace*{1.00cm}
\section{CONSTITUENT QUARK MODEL AND ITS UPDATED RESULTS OF VECTOR QUARKONIUM
RESONANCES}
\label{sec:CQM}

Spontaneous chiral symmetry breaking of the QCD Lagrangian together with the
perturbative one-gluon exchange (OGE) and the nonperturbative confining interaction
are the main pieces of constituent quark models. Using this idea, Vijande {\it et
al.}~\cite{Vijande:2004he} developed a model of the quark-quark interaction which is
able to describe meson phenomenology from the light to the heavy quark sector.

In the heavy quark sector chiral symmetry is explicitly broken and Goldstone-boson
exchanges do not appear. Thus, OGE and confinement are the only interactions
remaining. The one-gluon exchange potential is given by
\begin{widetext}
\begin{equation}
\begin{split}
&
V_{\rm OGE}^{\rm C}(\vec{r}_{ij}) =
\frac{1}{4}\alpha_{s}(\vec{\lambda}_{i}^{c}\cdot
\vec{\lambda}_{j}^{c})\left[ \frac{1}{r_{ij}}-\frac{1}{6m_{i}m_{j}} 
(\vec{\sigma}_{i}\cdot\vec{\sigma}_{j}) 
\frac{e^{-r_{ij}/r_{0}(\mu)}}{r_{ij}r_{0}^{2}(\mu)}\right], \\
& 
V_{\rm OGE}^{\rm T}(\vec{r}_{ij})=-\frac{1}{16}\frac{\alpha_{s}}{m_{i}m_{j}}
(\vec{\lambda}_{i}^{c}\cdot\vec{\lambda}_{j}^{c})\left[ 
\frac{1}{r_{ij}^{3}}-\frac{e^{-r_{ij}/r_{g}(\mu)}}{r_{ij}}\left( 
\frac{1}{r_{ij}^{2}}+\frac{1}{3r_{g}^{2}(\mu)}+\frac{1}{r_{ij}r_{g}(\mu)}\right)
\right]S_{ij}, \\
&
\begin{split}
V_{\rm OGE}^{\rm SO}(\vec{r}_{ij})= &  
-\frac{1}{16}\frac{\alpha_{s}}{m_{i}^{2}m_{j}^{2}}(\vec{\lambda}_{i}^{c} \cdot
\vec{\lambda}_{j}^{c})\left[\frac{1}{r_{ij}^{3}}-\frac{e^{-r_{ij}/r_{g}(\mu)}}
{r_{ij}^{3}} \left(1+\frac{r_{ij}}{r_{g}(\mu)}\right)\right] \times \\ & \times 
\left[((m_{i}+m_{j})^{2}+2m_{i}m_{j})(\vec{S}_{+}\cdot\vec{L})+
(m_{j}^{2}-m_{i}^{2}) (\vec{S}_{-}\cdot\vec{L}) \right].
\end{split}
\end{split}
\end{equation}
\end{widetext}

One characteristic of the model is the use of a screened linear confinement
potential. This has been able to reproduce the degeneracy pattern observed for the
higher excited states of light mesons~\cite{Segovia:2008zza}. As we assume that
confining interaction is flavor independent, we hope that this form of the potential
will be useful in our case because we are focusing on the high energy region of the
vector charmonium spectrum.

The different pieces of the confinement potential are
\begin{equation}
\begin{split}
&
V_{\rm CON}^{\rm C}(\vec{r}_{ij})=\left[-a_{c}(1-e^{-\mu_{c}r_{ij}})+\Delta
\right] (\vec{\lambda}_{i}^{c}\cdot\vec{\lambda}_{j}^{c}), \\
&
\begin{split}
&
V_{\rm CON}^{\rm SO}(\vec{r}_{ij}) =
-(\vec{\lambda}_{i}^{c}\cdot\vec{\lambda}_{j}^{c}) \frac{a_{c}\mu_{c}e^{-\mu_{c}
r_{ij}}}{4m_{i}^{2}m_{j}^{2}r_{ij}} \times \\
&
\times \left[((m_{i}^{2}+m_{j}^{2})(1-2a_{s})\right. \\
&
\quad\,\, +4m_{i}m_{j}(1-a_{s}))(\vec{S}_{+} \cdot\vec{L}) \\
&
\left. \quad\,\, +(m_{j}^{2}-m_{i}^{2}) (1-2a_{s}) (\vec{S}_{-}\cdot\vec{L})
\right].
\end{split}
\end{split}
\end{equation}

Further details about the quark model and the fine-tuned model parameters can be
found in Ref.~\cite{Segovia:2008zz}, where an attempt to describe the different
properties of vector charmonium resonances was done.

In Tables~\ref{tab:Jpsi} and~\ref{tab:upsilon} we summarize the model results for
vector charmonium and bottomonium states, respectively. We also compare with the
existing experimental data. In the bottomonium sector, only $S$-wave states are shown
because: i) $1^{--}$ $D$-wave states have not yet been observed; ii) $S\!\!-\!\!D$
splittings are expected to be very small in the bottomonium sector; and iii) they
are not very relevant in the course of this article.

As one can see in the case of $J^{PC}=1^{--}$ $c\bar{c}$ states, the agreement with
the experimental data is remarkable except for one state: the $X(4260)$ which do not
fit in the $Q\bar{Q}$ scheme.

\begin{table*}[!t]
\begin{center}
\begin{tabular}{cc|cc|ccc|ccc}
\hline
\hline
(nL) & States & ${\rm M}_{\rm The.}$ & ${\rm M}_{\rm Exp.}$ & $\Gamma^{e^{+}e^{-}}_{\rm The.} $ & $\Gamma^{e^{+}e^{-}}_{\rm Exp.}$ & & $\Gamma^{\rm S}_{\rm The.}$ & $\Gamma_{\rm Exp.}$ & \\ 
& & (MeV) & (MeV) & (keV) & (keV) & & (MeV) & (MeV) & \\
\hline
$(1S)$ & $J/\psi$     & $3096$ & $3096.916\pm0.011$       & $3.93$ & $5.55\pm0.14$ & & - & - & \\
$(2S)$ & $\psi(2S)$   & $3703$ & $3686.09\pm0.04$         & $1.78$ & $2.43\pm0.05$ & & - & - & \\
$(1D)$ & $\psi(3770)$ & $3796$ & $3772\pm1.1$             & $0.22$ & $0.22\pm0.05$ & & $26.5$ & $27.6\pm1.0$ & \\
%& $G(3900)$ & - & $3943\pm21$ & - & - & & - & - & \\
%& $X(4008)$ & - & $4008\pm40$ & - & - & & - & - & \\
$(3S)$ & $\psi(4040)$ & $4097$ & $4039\pm1$               & $1.11$ & $0.83\pm0.20$ & & $111.3$ & $80\pm10$ & \\
$(2D)$ & $\psi(4160)$ & $4153$ & $4153\pm3$               & $0.30$ & $0.48\pm0.22$ & & $116.0$ & $103\pm8$ & \\
       & $X(4260)$    & - & $4260\pm10$                   & -      & -             & & -       & -         & \\
$(4S)$ & $X(4360)$    & $4389$ & $4361\pm9$               & $0.78$ & -             & & $113.9$ & $74\pm18$ & \\
$(3D)$ & $\psi(4415)$ & $4426$ & $4421\pm4$               & $0.33$ & $0.58\pm0.07$ & $0.35\pm0.12$ & $159.0$ & $62\pm20$ & $119\pm16$ \\
$(5S)$ & $X(4640)$    & $4614$ & $4634^{+8+5}_{-7-8}$     & $0.57$ & -             & & $206.4$ & $92\pm52$ & \\
$(4D)$ & $X(4660)$    & $4641$ & $4664\pm11\pm5$          & $0.31$ & -             & & $135.1$ & $48\pm15$ & \\
\hline
\hline
\end{tabular}
\caption{\label{tab:Jpsi} Model results for $J^{PC}=1^{--}$ $c\bar{c}$ states
compared with the experimental data reported in PDG~\cite{Agashe:2014kda}.}
\end{center}
\end{table*}

\begin{table*}[!t]
\begin{center}
\begin{tabular}{cc|cc|cc|cc}
\hline
\hline
(nL) & States & ${\rm M}_{\rm The.}$ & ${\rm M}_{\rm Exp.}$ & $\Gamma^{e^{+}e^{-}}_{\rm The.} $ & $\Gamma^{e^{+}e^{-}}_{\rm Exp.}$ &  $\Gamma^{\rm S}_{\rm The.}$ & $\Gamma_{\rm Exp.}$ \\ 
&  & (MeV) & (MeV) & (keV) & (keV) & (MeV) & (MeV) \\
\hline
$(1S)$ & $\Upsilon(1S)$    & $9502$  & $9460\pm0.26$    & $0.71$ & $1.34\pm0.018$  & - & - \\
$(2S)$ & $\Upsilon(2S)$    & $10015$ & $10023.2\pm0.31$ & $0.37$ & $0.612\pm0.011$ & - & - \\
$(3S)$ & $\Upsilon(3S)$    & $10349$ & $10355.2\pm0.5$  & $0.27$ & $0.443\pm0.008$ & - & - \\
$(4S)$ & $\Upsilon(4S)$    & $10607$ & $10579.4\pm1.2$  & $0.21$ & $0.272\pm0.029$ & $20.59$ & $20.5\pm2.5$ \\
$(5S)$ & $\Upsilon(10860)$ & $10818$ & $10876\pm 11$    & $0.18$ & $0.31\pm0.07$   & $27.89$ & $55\pm28$  \\
$(6S)$ & $\Upsilon(11020)$ & $10995$ & $11019\pm8$      & $0.15$ & $0.130\pm0.030$ & $79.16$ & $79\pm16$  \\
\hline
\hline
\end{tabular}
\caption{\label{tab:upsilon} Model results  for
$J^{PC}=1^{--}$ $S$-wave $b\bar{b}$ states compared with the experimental data
reported in PDG~\cite{Agashe:2014kda}.}
\end{center}
\end{table*}

Usually the $\psi(4415)$ state has been assigned as a $4S$ state. Our particular
choice of the potential includes the new $X(4360)$ as a $4S$ state between the well
established $\psi(4160)$ and $\psi(4415)$ which are both predicted as $D$-wave
states. Whether or not this assignment is correct can be tested with the $e^{+}e^{-}$
leptonic widths. From Table~\ref{tab:Jpsi} one can see that the width of
the $4S$ state is $0.78\,{\rm keV}$, whereas the last experimental value for the
$\psi(4415)$ is $\Gamma_{e^{+}e^{-}}=0.35\pm0.12\,{\rm keV}$, in excellent agreement
with the result for the $3D$ state ($0.33\,{\rm keV}$). The measurement of the
leptonic width for the $X(4360)$ is very important and will clarify the situation. 

Furthermore, it has been shown in Ref.~\cite{Segovia:2011zza} that the assignment of
the $X(4360)$ and $\psi(4415)$ resonances as the $4S$ and $3D$ $J^{PC}=1^{--}$
$c\bar{c}$ states, respectively, is compatible with the data of the exclusive
reactions $e^{+}e^{-}\to D^{0}D^{-}\pi^{+}$ and $e^{+}e^{-}\to
D^{0}D^{\ast-}\pi^{+}$.

The total decay widths shown in Table~\ref{tab:Jpsi} differ with respect to those
calculated in Ref.~\cite{Segovia:2008zz}. The reason for that is the following. The
$^{3}P_{0}$ model presents a parameter, the strength $\gamma$ of the decay
interaction, which is regarded as a free flavor independent constant and is fitted to
the data. In Ref.~\cite{Segovia:2008zz} the parameter $\gamma$ of the $^{3}P_{0}$
model was adjusted to reproduce the decay rate $\psi(3770)\to D^{+}D^{-}$, which is
the only decay rate well established in the vector charmonium sector. However, in
Ref.~\cite{Segovia:2012cd}, we performed a calculation of the strong decay widths of
the mesons which belong to charmed, charmed-strange, hidden charm and hidden bottom
sectors. A global fit of the experimental data shows that, contrarily to the usual
wisdom, the $\gamma$ depends on the flavor of the quark-antiquark in the decaying
meson. We proposed a $\gamma$ strength parameter that depends on the scale through
the reduced mass of the $q\bar{q}$ pair in the decaying meson. The results predicted
by the $^{3}P_{0}$ model with the suggested running of the $\gamma$ parameter are in
a global agreement with the experimental data, being remarkable in most of the cases
studied.

Above the $\psi(4415)$ our model predicts two states which have been assigned as
the $X(4640)$ and $X(4660)$ despite of the total decay widths are not so well
reproduce. The lower value of the total width of the $X(4660)$ favors the $4D$
assignment for this state although interference between the two states can be the
origin of the poor description of the total widths. A similar situation has been
recently observed by the LHCb experiment~\cite{Aaij:2014xza} in which the
$D_{s1}(2860)$ signal has resulted in a couple of resonances that compite and obscure
the experimental measurement of the total decay width.

The model predicts reasonably well the masses and total decay widths of the
$b\bar{b}$ states. Our results for the leptonic widths are suppressed. It is
noteworthy that the experimental value of the leptonic width of the $\Upsilon(10860)$
does not follow the pattern elucidated by the other resonances.

%%%%%%%%%%%%%%%%%%%%%%%%%%%%%%%%%%%%%%%%%%%%%%%%%%%%%%%%%%%%%%%%%%%%%%%%%%%%%%%%

\section{QCD MULTIPOLE EXPANSION}
\label{sec:QCDME}

Hadronic transitions between color singlet states involve, at least, the emission of
two gluons. These gluons are rather soft because the energy difference between the
initial and final quarkonium state are usually small.
Gottfried~\cite{Gottfried:1977gp} has pointed out that this gluon radiation can be
treated in a multipole expansion since the wavelengths of the emitted gluons are
large compared with the size of the $Q\bar{Q}$ states. After the expansion of the
gluon field, the Hamiltonian of the system can be decomposed as
\begin{equation}
{\cal H}^{\rm eff}_{\rm QCD} = {\cal H}^{0}_{\rm QCD} + {\cal H}^{1}_{\rm QCD} +
{\cal H}^{2}_{\rm QCD},
\label{eq:Hqcd}
\end{equation}
with ${\cal H}^{0}_{\rm QCD}$ the sum of the kinetic and potential energies of the
heavy quarks, and ${\cal H}^{1}_{\rm QCD}$ and ${\cal H}^{2}_{\rm QCD}$ defined by 
\begin{equation}
\begin{split}
{\cal H}^{1}_{\rm QCD} &= Q_{a} A^{a}_{0}(x,t), \\
{\cal H}^{2}_{\rm QCD} &=-d_{a} E^{a}(x,t) - m_{a} B^{a}(x,t),
\label{eq:Hqcd2}
\end{split}
\end{equation}
in which $Q_{a}$, $d_{a}$ and $m_{a}$ are the color charge, the color electric dipole
moment and the color magnetic dipole moment, respectively. As the $Q\bar{Q}$ is a
color singlet, there is no contribution of the ${\cal H}^{1}_{\rm QCD}$ and only
$E_{l}$ and $B_{m}$ transitions occur. The lowest order term between two color
singlets involves two gluons and therefore the lowest multipole is $E_{1}E_{1}$.

The gauge-invariant formulation of multipole expansion within QCD was given by
Tung-Mow Yan in Ref.~\cite{Yan:1980uh}. We will follow the updated
review~\cite{Kuang:2006me} and references therein to calculate the hadronic
transitions in which we are interested.

\subsection{Two pion hadronic transitions}
\label{subsec:Spinnonflip}

These processes are dominated by double electric-dipole transitions (E1E1). The
transition amplitude is given by~\cite{Kuang:2006me}
\begin{equation}
{\cal M}_{E1E1}=i\frac{g_{E}^{2}}{6} \left\langle\right.\!\! \Phi_{F}h \,
|\vec{x}\cdot\vec{E} \, \frac{1}{E_{I}-H^{(0)}_{QCD}-iD_{0}} \,
\vec{x}\cdot\vec{E}| \, \Phi_{I} \!\! \left.\right\rangle,
\label{eq:E1E1}
\end{equation}
where $\vec{x}$ is the separation between $Q$ and $\bar{Q}$, and
$(D_0)_{bc}\equiv\delta_{bc}\partial_{0}-g_{s}f_{abc}A^{a}_{0}$.

Inserting a complete set of intermediate states the transition
amplitude~(\ref{eq:E1E1}) becomes
\begin{equation}
{\cal M}_{E1E1}=i\frac{g_{E}^{2}}{6} \sum_{KL}
\frac{\left\langle\right.\!\! \Phi_{F}|x_k|KL \!\!\left.\right\rangle
\left\langle\right.\!\! KL|x_l|\Phi_I \!\!\left.\right\rangle}{E_I-E_{KL}}
\left\langle\right.\!\! \pi\pi|E^{a}_{k} E^{a}_{l}|0 \!\!\left.\right\rangle,
\label{eq:factorizedE1E1}
\end{equation}
where $E_{KL}$ is the energy eigenvalue of the intermediate state $|KL\rangle$ with
the principal quantum number $K$ and the orbital angular momentum $L$.

The intermediate states in the hadronic transition are those produced after the
emission of the first gluon and before the emission of the second gluon. They are
states with a gluon and a color-octet $Q\bar{Q}$ and thus these states are the
so-called hybrid states. It is difficult to calculate these hybrid states from first
principles of QCD. So we take a reasonable model, which will be explained below, to
describe them.

The transition amplitude~(\ref{eq:factorizedE1E1}) split into two factors. The first
one concerns to the wave functions and energies of the initial and final quarkonium
states as well as those of the intermediate hybrid mesons. All these quantities can
be calculated using suitable quark models. The second one describes the conversion of
the emitted gluons into light hadrons. As the momenta involved are very low this
matrix element cannot be calculated using perturbative QCD and one needs to resort to
a phenomenological approach based on soft-pion techniques~\cite{Brown:1975dz}. In the
center-of-mass frame, the two pion momenta $q_{1}$ and $q_{2}$ are the only
independent variables describing this matrix element which, in the nonrelativistic
limit, can be parametrized as~\cite{Brown:1975dz,Yan:1980uh,Kuang:2006me}
\begin{equation}
\begin{split}
& \frac{g_{E}^{2}}{6} \left\langle\right.\!\!
\pi_{\alpha}(q_{1})\pi_{\beta}(q_{2})|E^{a}_{k}E^{a}_{l}|0
\!\!\left.\right\rangle =
\frac{\delta_{\alpha\beta}}{\sqrt{(2\omega_{1})(2\omega_ {2})}} \,\times \\
&
\times
\left[C_{1}\delta_{kl}q^{\mu}_{1}q_{2\mu} + C_{2}\left(q_{1k}q_{2l}+q_{1l}q_{2k}
-\frac{2}{3}\delta_{kl}\vec{q}_{1}\cdot\vec{q}_{2}\right)\right],
\label{HofE1E1}
\end{split}
\end{equation}
where $C_{1}$ and $C_{2}$ are two unknown constants. 

Finally, the transition rate is given by~\cite{Kuang:1981se}
\begin{widetext}
\vspace*{-0.60cm}
\begin{equation}
\begin{split}
\Gamma(\Phi_{I}(^{2s+1}{l_{I}}_{J_{I}}) \to 
\Phi_{F}(^{2s+1}{l_{F}}_{J_{F}})\pi\pi) =&
\delta_{l_{I}l_{F}}\delta_{J_{I}J_{F}} (G|C_{1}|^{2}-\frac{2}{3}H|C_{2}|^{2}
)\left|\sum_{L}(2L+1) \left(\begin{matrix} l_{I} & 1 & L \\ 0 & 0 & 0
\end{matrix}\right) \left(\begin{matrix} L & 1 & l_{I} \\ 0 & 0 & 0
\end{matrix}\right) f_{IF}^{L}\right|^{2} \\
&
+(2l_{I}+1)(2l_{F}+1)(2J_{F}+1) \sum_{k} (2k+1) (1+(-1)^{k})
\left\lbrace\begin{matrix} s & l_{F} & J_{F} \\ k & J_{I} & l_{I}
\end{matrix}\right\rbrace^{2} H |C_{2}|^{2} \times \\
&
\times\left|\sum_{L} (2L+1) \left(\begin{matrix} l_{F} & 1 & L \\ 0 & 0 & 0
\end{matrix}\right) \left(\begin{matrix} L & 1 & l_{I} \\ 0 & 0 & 0
\end{matrix}\right) \left\lbrace\begin{matrix} l_{I} & L & 1 \\ 1 & k & l_{F}
\end{matrix}\right\rbrace f_{IF}^{L} \right|^{2},
\label{eq:gamapipi}
\end{split}
\end{equation}
with
\begin{equation}
f_{IF}^{L}=\sum_{K}\frac{1}{M_{I}-M_{KL}}\left[\int dr r^{3}
R_{F}(r)R_{KL}(r)\right] \left[\int dr' r'^{3} R_{KL}(r') R_{I}(r')\right],
\label{eq:fifl}
\end{equation}
\end{widetext}
where $R_{I}(r)$, $R_{F}(r)$ and $R_{KL}(r)$ are the radial wave functions of the
initial, final and intermediate vibrational states, respectively. $M_{I}$ is
the mass of the decaying meson and $M_{KL}$ are the masses of the intermediate
vibrational states. The quantities $G$ and $H$ are the phase-space integrals
\vspace*{-0.30cm}
\begin{equation}
\begin{split}
G=&\frac{3}{4}\frac{M_{F}}{M_{I}}\pi^{3}\int
dM_{\pi\pi}^{2}\,K\,\left(1-\frac{4m_{\pi}^{2}}{M_{\pi\pi}^{2}}\right)^{1/2}(M_{
\pi\pi}^{2}-2m_{\pi}^{2})^{2}, \\
H=&\frac{1}{20}\frac{M_{F}}{M_{I}}\pi^{3}\int
dM_{\pi\pi}^{2}\,K\,\left(1-\frac{4m_{\pi}^{2}}{M_{\pi\pi}^{2}}\right)^{1/2}
\times \\
&
\times\left[(M_{\pi\pi}^{2}-4m_{\pi}^{2})^{2}\left(1+\frac{2}{3}\frac{K^{2}}{M_{
\pi\pi}^{2}}\right)\right. \\
&
\left.\quad\,\, +\frac{8K^{4}}{15M_{\pi\pi}^{4}}(M_{\pi\pi}^{4}+2m_{\pi}^{2}
M_{\pi\pi}^{2}+6m_{\pi}^{4})\right],
\end{split}
\end{equation}
with $K$ given by
\begin{equation}
K = \frac{\sqrt{\left[(M_{I}+M_{F})^{2}-M_{\pi\pi}^{2}\right]
\left[(M_{I}-M_{F})^{2}-M_{ \pi\pi}^{2}\right]}}{2M_{I}}.
\end{equation}

\subsection{A model for hybrid mesons}
\label{subsec:hybrids}

From the generic properties of QCD, we might expect to have states in which the
gluonic field itself is excited and carries $J^{PC}$ quantum numbers. A bound-state
is called glueball when any valence quark content is absent, the addition of a
constituent quark-antiquark pair to an excited gluonic field gives rise to what is
called a hybrid meson. The gluonic quantum numbers couple to those of the $q\bar{q}$
pair. This coupling may give rise to so-called exotic $J^{PC}$ mesons, but also can
produce hybrid mesons with natural quantum numbers. We are interested on the last
ones because they are involved in the calculation of hadronic transitions within the
QCDME approach.

As stated in the introduction, estimates of hybrid meson properties have
traditionally followed from different models. Among them, we adopt the QCS model
since it was used in the early works of QCDME and it incorporates finite quark mass
corrections. The QCS model is defined by a relativistic-, gauge- and
reparametrization-invariant action describing quarks interacting with color $SU(3)$
gauge fields in a two dimensional world sheet. It is assumed that the meson is
composed of a quark and antiquark linked by an appropriate color electric flux line
(the string).

The string can carry energy-momentum only in the region between the quark and the
antiquark. The string and the quark-antiquark pair can rotate as a unit and also
vibrate. Ignoring its vibrational motion, the equation which describes the dynamics
of the quark-antiquark pair linked by the string should be the usual Schr\"odinger
equation with a confinement potential. Gluon excitation effects are described by the
vibration of the string. These vibrational modes provide new states beyond the naive
meson picture.

A complete description of the model can be found
in Refs.~\cite{Tye:1975fz,Giles:1977mp,Buchmuller:1979gy}. We will give here only a
brief description of it. The dynamics of the string is defined by the action
\begin{equation}
\begin{split}
S &= \int^{\infty}_{-\infty} \, d^2u \, \sqrt{-g} \, \times \\
&
\times \bigg\{ \sum_j\bar{\psi}_j \left[\gamma_\mu
\tau^{\alpha\mu}\left(\frac{i}{2}\bar{\partial_{\alpha}}
-eB_{a\alpha}T^{\alpha}\right)-M_j\right] \psi_j \\
&
\hspace{0.50cm} -\frac{1}{4}F_{a\alpha\beta}F^{\alpha\beta}_a \bigg\},
\end{split}
\end{equation}
where $\psi_j(u)$ is a four-component Dirac field, $d^2u\sqrt{-g}$ is the invariant
volume element, $T^a$ are the eight matrix generators of $SU(3)$ color and
$B_{a\alpha}$ are de color gauge fields. From this action, in the nonrelativistic
limit, one obtains the effective Hamiltonian~\cite{Giles:1977mp} composed of three
terms (the quark, the string and the Coulomb):
\begin{equation}
\begin{split}
{\cal H} &= {\cal H}_q + {\cal H}_s + {\cal H}_c \\
&
=\int d\sigma\chi^+\left(M\beta-i\alpha_1 \partial_1\right)\chi \\
&
\hspace{0.50cm} +\int d\sigma \chi^+\beta \chi\frac{Mv^2}{2} \\
&
\hspace{0.50cm} +\frac{e^2}{2}\int d\sigma
d\sigma'\chi^+(\sigma)T^a\chi(\sigma)G(\sigma,\sigma')\chi^+(\sigma')
T^a\chi(\sigma'),
\end{split}
\end{equation}
which, in absence of vibrations and after quantization of the rotational modes, leads
to the following Schr\"odinger equation for the meson bound-states in the
center-of-mass frame
\begin{equation}
\left[2M-\frac{1}{M}\frac{\partial^2}{\partial r^2}+kr-\frac{l(l+1)}{Mr^2}\right]
\psi(r) = E\psi(r).
\end{equation}

The coupled equations that describe the dynamics of the string and the quark
sectors are very nonlinear so that there is no hope of solving them completely. Then,
to introduce the vibrational modes, we use the following approximation scheme. First,
we solve the string Hamiltonian (via de Bohr-Oppenheimer method) to obtain the
vibrational energies as functions of $r$, the interquark distance. These are then
inserted into the meson equation as an effective potential, $V_{n}(r)$, that depends
on the distance between the quark and the antiquark.

Assuming the quark mass to be very heavy so that the ends of the string are fixed,
the vibrational potential energy can be estimated using the Bohr-Sommerfeld
quantization to be~\cite{Giles:1977mp}
\begin{equation}
\begin{split}
V_{n}(r) = \sigma r\left\lbrace 1 + \frac{2n\pi}{\sigma
\left[(r-2d)^{2}+4d^{2}\right]} \right\rbrace^{1/2},
\end{split}
\end{equation}
where $d$ is the correction due to the finite quark mass 
\begin{equation}
d(m_{Q},r,\sigma,n)=\frac{\sigma r^{2}\alpha_{n}}{4(2m_{Q} + \sigma
r\alpha_{n})},
\end{equation}
being $\alpha_{n}$ a parameter related with the shape of the vibrating
string~\cite{Giles:1977mp}, and can take the values $1\leq\alpha_{n}^{2}\leq2$. For
$n=0$, $V_{n}(r)$ reduces to the naive $Q\bar{Q}$ one. 

In our quark model, the central part of the confining potential has the following
form
\begin{equation}
V_{\rm CON}^{\rm C}(r) = \frac{16}{3} [a_{c}(1-e^{-\mu_{c}r})-\Delta],
\end{equation}
and can be written as
\begin{equation}
V_{\rm CON}^{\rm C}(r) = \sigma(r)r + \mbox{cte,}
\end{equation}
where
\begin{equation}
\begin{split}
\sigma(r) &= \frac{16}{3}\,a_{c}\,\left(\frac{1-e^{-\mu_{c}r}}{r}\right), \\
\mbox{cte} &= -\frac{16}{3}\,\Delta.
\end{split}
\end{equation}
This means that our effective string tension, $\sigma(r)$, is not a constant but
depends on the interquark distance, $r$. In fact, it decreases with respect to $r$
until it reaches the string breaking region.

Following the ideas of Ref.~\cite{Buchmuller:1979gy}, the potential for hybrid mesons
derived from our constituent quark model has the following expression
\begin{equation}
V_{\rm hyb}(r)=V_{\rm OGE}^{\rm C}(r) + V_{\rm CON}^{\rm C}(r) +
\left[V_{n}(r) - \sigma(r)r\right]  ,
\label{eq:pothyb}
\end{equation}
where we have not taken into account the spin-dependent terms. $V_{\rm OGE}^{\rm
C}(r) + V_{\rm CON}^{\rm C}(r)$ is the naive quark-antiquark potential and $V_{n}(r)$
is the vibrational one. We must subtract the term $\sigma(r)r$ because it appears
twice, one in $V_{\rm CON}^{\rm C}(r)$ and the other one in $V_{n}(r)$. This
potential does not include new parameters besides those of the original quark model. 
In that sense the calculation of the hybrids states is parameter-free. More
explicitly, our different contributions are
\begin{equation}
\begin{split}
V_{\rm OGE}^{\rm C}(r) &= -\frac{4\alpha_{s}}{3r}, \\
V_{\rm CON}^{\rm C}(r) &= \frac{16}{3} [a_{c}(1-e^{-\mu_{c}r})-\Delta], \\
V_{n}(r) &= \sigma(r)r \left\lbrace 1 + \frac{2n\pi}{\sigma(r)
\left[(r-2d)^{2}+4d^{2}\right]} \right\rbrace^{1/2},
\end{split}
\end{equation}
where
\begin{equation}
d(m_{Q},r,\sigma,n) =
\frac{\sigma(r)r^{2}\alpha_{n}}{4(2m_{Q}+\sigma(r)r\alpha_{n})}.
\end{equation}
One can realize that, just like the naive quark model, the hybrid potential has a
threshold defined by
\begin{equation}
V_{\rm hyb}(r) \xrightarrow{r\to\infty} \frac{16}{3} \left(a_{c}-\Delta\right).
\end{equation}

%%%%%%%%%%%%%%%%%%%%%%%%%%%%%%%%%%%%%%%%%%%%%%%%%%%%%%%%%%%%%%%%%%%%%%%%%%%%%%%%

\section{RESULTS}
\label{sec:results}

All the parameters of the quark model are taken from~\cite{Segovia:2008zz} then we
only need to fix the two parameters $C_{1}$ and $C_{2}$ of Eq.~(\ref{HofE1E1}).

For a given $M_{\pi\pi}$ invariant mass $C_{1}$ term is isotropic and therefore
contributes to the $S$-state into $S$-state transitions while the $C_{2}$ term has a
$L=2$ angular dependence and contributes to the $D$-wave into $S$-wave transitions.
Then, we can use the well established $\psi(2S)\to J/\psi\pi^{+}\pi^{-}$ and
$\psi(3770)\to J/\psi\pi^{+}\pi^{-}$ transitions to fix the two parameters. It is
argued sometimes that to reproduce the leptonic decay width of the $\psi(3770)$,
this state should be a mixture like $\psi(3770) = \left|2S\right\rangle
\sin\vartheta + \left|1D\right\rangle \cos\vartheta$ being $\vartheta$ an adjustable
parameter. In our model the angle $\vartheta$ is determined by the dynamics to be
$\vartheta\approx1^{\circ}$ although a correct value of the leptonic width is
obtained (see Table~\ref{tab:Jpsi}). Therefore, we will consider the $\psi(3770)$ as
a pure $D$-wave state. Taken the experimental values of the widths of these two
resonances from PDG~\cite{Agashe:2014kda} we obtain
\begin{equation}
\begin{split}
|C_{1}|^{2} &= (9.396\pm0.503)\times10^{-5}, \\
|C_{2}|^{2} &= (3.051\pm0.430)\times10^{-4}.
\end{split}
\end{equation}
The value of the ratio $C_2/C_1\!\sim\!1.8$ is compatible with the
literature~\cite{Brambilla:2010cs}.

From Eq.~(\ref{eq:factorizedE1E1}) one can see that the quantum numbers of hybrid
states which participate in the two pion transitions are $J^{PC}=1^{--}$. In
Table~\ref{tab:CQMccg} we show the values of the masses of the hybrid states with
these quantum numbers and different radial excitations in the charmonium and
bottomonium sector.

\begin{table}[!t]
\begin{center}
\begin{tabular}{ccc}
\hline
\hline
\tstrut
$K$ & $c\bar cg$ & $b\bar bg$ \\
\hline
\tstrut
$1$ & $4351$ & $10785$ \\
$2$ & $4639$ & $10999$ \\
$3$ & $4855$ & $11175$ \\
$4$ & $5020$ & $11325$ \\
$5$ & $5145$ & $11452$ \\
$6$ & $5238$ & $11562$ \\
\hline
\hline
\end{tabular}
\caption{\label{tab:CQMccg} $J^{PC}=1^{--}$ hybrid meson masses, in MeV, calculated
in the $c\bar{c}$ and $b\bar{b}$ sectors. The variation of the parameter $\alpha_{n}$
which range between $1<\alpha_{n}<\sqrt{2}$ modifies the energy as much as $30\,{\rm
MeV}$, we have taken $\alpha_{n}=\sqrt{1.5}$.}
\end{center}
\end{table}

The mass of the ground state in the $c\bar{c}g$ sector is $4.35\,{\rm GeV}$ which
agrees roughly with the results of the flux-tube model~\cite{Barnes:1995hc}
$(4.1-4.2)$, Coulomb gauge QCD~\cite{Guo:2008yz} $(4.47)$, QCD string
model~\cite{Kalashnikova:2008qr} $(4.397)$, potential model~\cite{Ke:2007ih}
$(4.23)$ and lattice calculation~\cite{Dudek:2008sz} $(4.40)$.

Typically, in experiments that cover hadronic transitions of heavy quarkonia with
two-pion emission, the quantities which are measured are the cross section and the
product ${\cal B}_{\pi^+\pi^-(Q\bar{Q})}\times \Gamma_{e^{+}e^{-}}$ where ${\cal
B}_{\pi^+\pi^-(Q\bar{Q})}$ indicates the branching ratio of the decay and
$\Gamma_{e^{+}e^{-}}$ is the leptonic width of the resonance. We will give results
for the product ${\cal B}_{\pi^+\pi^-(Q\bar{Q})}\times \Gamma_{e^{+}e^{-}}$ and
values of the cross section at peak.
  
\begin{table}[!t]
\begin{center}
\begin{tabular}{c|cc|cc}
\hline
\hline
\tstrut
Initial Meson & $R^{\rm th}_{J/\psi}$ (eV) & $R^{\rm ex}_{J/\psi}$ (eV) &
$R^{\rm th}_{\psi(2S)}$ (eV) & $R^{\rm ex}_{\psi(2S)}$ (eV) \\
\hline
$\psi(4040)$ & $1.20$            & - & $0.11$            & - \\
$\psi(4160)$ & $0.40$            & - & $6\times 10^{-2}$ & - \\
$X(4360)$    & $52.5$            & - & $5.05$            & $7.4\pm 0.9$ \\
$X(4415)$    & $3\times 10^{-6}$ & - & $0.27$            & - \\
$X(4660)$    & $0.58$            & - & $1.08$            & $1.04\pm 0.5$ \\
\hline
\hline
\end{tabular}
\caption{\label{tab:pipicc} $R_{\psi(nS)} = {\cal B}_{\pi^{+}\pi^{-}\psi(nS)} \times
\Gamma_{e^{+}e^{-}}$ for the $J^{PC}= 1^{--}$ $S$-wave charmonium states.
Experimental data are from Ref.~\cite{Lees:2012pv}.}
\end{center}
\end{table}

\begin{table}[!t]
\begin{center}
\begin{tabular}{c|cc|cc}
\hline
\hline
\tstrut
Initial Meson & $\sigma^{\rm th}_{J/\psi}$ (pb) & $\sigma^{\rm ex}_{J/\psi}$ (pb) &
$\sigma^{\rm th}_{\psi(2S)}$ (pb) & $\sigma^{\rm ex}_{\psi(2S)}$ (pb) \\
\hline
$\psi(4040)$ & $13.46$           & - & $1.25$  & -         \\
$\psi(4160)$ & $3.32$            & - & $0.50$  & -         \\
$X(4360)$    & $329.7$           & - & $31.69$ & $52\pm 2$ \\
$X(4415)$    & $4\times 10^{-5}$ & - & $3.38$  & -         \\
$X(4660)$    & $10.97$           & - & $20.29$ & $28\pm 2$ \\
\hline
\hline
\end{tabular}
\caption{\label{tab:cscc} The cross section at peak for the $J^{PC}= 1^{--}$ $S$-wave
charmonium states. Experimental data are from Ref.~\cite{Lees:2012pv}.}
\end{center}
\end{table}

Table~\ref{tab:pipicc} shows the calculated ${\cal B}_{\pi^{+}\pi^{-}\psi(nS)} \times
\Gamma_{e^{+}e^{-}}$ for the $J^{PC}= 1^{--}$ charmonium states. As the decays
$\psi(2S)\to J/\psi \pi^{+}\pi^{-}$ and $\psi(3770)\to J/\psi \pi^{+}\pi^{-}$ have
been used to fit the $C_{1}$ and $C_{2}$ parameters they are not included in the
table. One can see that in the case of the decay channel $\psi(2S)\pi^{+}\pi^{-}$ the
only significant values correspond to the decays of the $X(4360)$ and $X(4660)$
which are also in agreement with the recent experimental data. These results justify
why no signals of the $\psi(4040)$, $\psi(4160)$ and $\psi(4415)$ have been seen in
the data. In the decay channel $J/\psi\pi^{+}\pi^{-}$ a high value of the
${\cal B}_{\pi^{+}\pi^{-}\psi(nS)}\times\Gamma_{e^{+}e^{-}}$ is obtained for the
$X(4360)$ resonance. This result apparently contradicts the experimental data because
this decay has not been reported in the reaction $e^{+}e^{-}\to
J/\psi\pi^{+}\pi^{-}$~\cite{Lees:2012cn}. The cross section of this reaction shows a
resonance in the $4.2-4.4$ energy region which has been attributed to the $X(4260)$,
which does not appear in our calculation as a $c\bar{c}$ meson. However, an
interference between the $X(4260)$ and $X(4360)$ resonances would be possible. The
values for the rest of the resonances are small and could be the reason for
not seen in the experiment.

Same conclusions can be obtained from the values of the cross section at peak
(Table~\ref{tab:cscc}). The two measured values in the $\psi(2S)\pi^{+}\pi^{-}$
channel are in agreement with our theoretical results and the only significant cross
section at peak is obtained for the $X(4360)$ resonance in the $J/\psi\pi^{+}\pi^{-}$
channel.

\begin{table*}[!t]
\begin{center}
\begin{tabular}{c|cc|cc|cc}
\hline
\hline
\tstrut
Initial Meson & $R^{\rm th}_{\Upsilon (1S)}$ (eV) & $R^{\rm ex}_{\Upsilon(1S)}$ (eV)
& $R^{\rm th}_{\Upsilon(2S)}$ (eV) & $R^{\rm ex}_{\Upsilon(2S)}$ (eV) &
$R^{\rm th}_{\Upsilon(3S)}$ (eV) & $R^{\rm ex}_{\Upsilon(3S)}$
(eV) \\
\hline
$\Upsilon(2S)$    & $98.34$ & $105.4\pm 4.3$ & -    & - & - & -\\
$\Upsilon(3S)$    & $23.94$ & $18.5\pm 9.8$  & 5.58 & -\\
$\Upsilon(4S)$    & $6\times 10^{-2}$        & $(2.3\pm 0.9)\times 10^{-2}$ & $2.5\times10^{-3}$  & $(2.3\pm 0.4)\times 10^{-2}$   & - & -\\
$\Upsilon(10860)$ & $4.1\times 10^{-2}$      & $1.64 \pm 0.40$ & $5.8 \times 10^{-2}$ & $ 2.42\pm 0.64$ & $1.8\times 10^{-2} $ & $1.49\pm 0.65$ \\
\hline
\hline
\end{tabular}
\caption{\label{tab:pipibb} $R_{\Upsilon(nS)} = {\cal B}_{\pi^{+}\pi^{-}\Upsilon(nS)}
\times \Gamma_{e^{+}e^{-}}$ for the $J^{PC}= 1^{--}$ $S$-wave bottomonium states.
Experimental data are  from Ref.~\cite{Agashe:2014kda}.}
\end{center}
\end{table*}

The results for the bottomonium sector are shown in Tables~\ref{tab:pipibb}
and~\ref{tab:csbb}. One can see that the theoretical values agree reasonably well 
with the experimental ones except in the case of the $\Upsilon(10860)$. We do not
find any hybrid state around the $\Upsilon(10860)$ mass region. Therefore, the
mechanism which explains the large widths in the charm sector cannot be applied to
this case. It seems that the anomalous width of the $\Upsilon(10860)$ can be
justified including tetraquark components in its wave function~\cite{Ali:2009es}.

\begin{table*}[!t]
\begin{center}
\begin{tabular}{c|cc|cc|cc}
\hline
\hline
\tstrut
Initial Meson & $\sigma^{\rm th}_{\Upsilon (1S)}$ (pb) & $\sigma^{\rm ex}_{\Upsilon(1S)}$ (pb) & $\sigma^{\rm th}_{\Upsilon(2S)}$
(pb) & $\sigma^{\rm ex}_{\Upsilon(2S)}$ (pb) & $\sigma^{\rm th}_{\Upsilon(3S)}$
(pb) & $\sigma^{\rm ex}_{\Upsilon(3S)}$ (pb)\\
\hline
$\Upsilon(2S)$    & $4.49 \times 10^5$   & - & - & - & - & - \\
$\Upsilon(3S)$    & $1.61 \times 10^5$   & - & $0.38\times 10^5$    & - & - & - \\
$\Upsilon(4S)$    & $0.39$               & - & $1.57\times 10^{-2}$ & - & - & - \\
$\Upsilon(10860)$ & $9.38\times 10^{-2}$ & $2.27\pm 0.14$ & $1.31\times 10^{-1}$ & $4.07\pm 0.45$ & $4.06\times 10^{-2}$ & $1.46\pm 0.16$ \\
\hline
\hline
\end{tabular}
\caption{\label{tab:csbb} The cross section at peak for the $J^{PC}= 1^{--}$ $S$-wave
bottomonium states. Experimental data are from Ref.~\cite{Agashe:2014kda}.}
\end{center}
\end{table*}

%%%%%%%%%%%%%%%%%%%%%%%%%%%%%%%%%%%%%%%%%%%%%%%%%%%%%%%%%%%%%%%%%%%%%%%%%%%%%%%%

\section{Summary}
\label{sec:summary}

Hadronic transitions of heavy quarkonia with two-pion emission have been calculated
in the framework of the QCD multipole expansion (QCDME). Charmonium and bottomonium
states are described in a constituent quark model whereas the hybrid intermediate
states, needed in the QCDME method, are calculated in a natural extension of the
constituent quark model. This extension is based on the quark confining string scheme
which does not include any new parameter.

We have analyzed the $J/\psi\pi^{+}\pi^{-}$ and $\psi(2S)\pi^{+}\pi^{-}$ channels for
charmonium decays and the $\Upsilon(nS)\pi^{+}\pi^{-}$ with $n=1$, $2$ and $3$ for
bottomonium decays.

In the invariant mass distribution process $e^{+}e^{-}\to \psi(2S)\pi^{+}\pi^{-}$ two
resonance structures appear at the masses of the $X(4360)$ and $X(4660)$ states. Our
calculation shows that the only significant transition rates correspond to these
resonances being the rest, at least, one order of magnitude smaller. This result is
explained by the presence of two hybrid states, each of them located near the masses
of the $X(4360)$ and $X(4660)$, which enhance the transition rates.

In the $e^{+}e^{-}\to J/\psi\pi^{+}\pi^{-}$ reaction, there is only a significant
peak in the mass region of the $X(4260)$. As most of the potential models, our model
is not able to describe this peculiar state which is only seen in this transition but
not in any open-charm decay channel as the rest of the $J^{PC}=1^{--}$ charmonium
states. Our theoretical results indicate that the only significant transition rate in
the $J/\psi\pi^{+}\pi^{-}$ channel corresponds to the $X(4360)$ resonance. This
tension between the theoretical and experimental results requires more accurate
studies.

In the bottomonium sector our results agree with the experimental data except in the
case of the $\Upsilon(10860)$ resonance. This may suggest a more complex structure
(tetraquark or molecule) for this state.

%%%%%%%%%%%%%%%%%%%%%%%%%%%%%%%%%%%%%%%%%%%%%%%%%%%%%%%%%%%%%%%%%%%%%%%%%%%%%%%%

\section{Acknowledgements}
\label{sec:acknowledgements}

This work has been partially funded by U.\,S.\,Department of Energy, Office of
Nuclear Physics, contract no.~DE-AC02-06CH11357, by Ministerio de Ciencia y
Tecnolog\'\i a under Contract no. FPA2010-21750-C02-02, by the European
Community-Research Infrastructure Integrating Activity ``Study of Strongly
Interacting Matter'' (HadronPhysics3 Grant no. 283286) and by the Spanish
Ingenio-Consolider 2010 Program CPAN (CSD2007-00042).

%%%%%%%%%%%%%%%%%%%%%%%%%%%%%%%%%%%%%%%%%%%%%%%%%%%%%%%%%%%%%%%%%%%%%%%%%%%%%%%%

% Including bibliography through bibtex

\bibliographystyle{apsrev}
\bibliography{hadronic_transitions_v7}

\end{document}